\documentclass[%
 reprint,
superscriptaddress,
 amsmath,amssymb,
 aps,
]{revtex4-2}
\usepackage{graphicx}% Include figure files
\usepackage{dcolumn}% Align table columns on decimal point
\usepackage{bm}% bold math
\usepackage{color}
\usepackage{soul}
\usepackage[hidelinks]{hyperref}
\usepackage{natbib}
\begin{document}

% Evolution of ultra-slow shock waves observed in a tunable magnetic lattice
\title{Observation of ultra-slow shock waves in a tunable magnetic lattice}

\author{Jian Li}
\affiliation{ Massachusetts Institute of Technology, Department of Civil and Environmental Engineering, Cambridge, MA, 02139, USA}
\author{S Chockalingam}
\affiliation{ Department of Aeronautics and Astronautics, Massachusetts Institute of Technology, Cambridge, MA 02139, USA}
\author{Tal Cohen}
\email{Corresponding author: talco@mit.edu}
\affiliation{ Massachusetts Institute of Technology, Department of Civil and Environmental Engineering, Cambridge, MA, 02139, USA}
\affiliation{ Massachusetts Institute of Technology, Department of Mechanical Engineering, Cambridge, MA, 02139, USA}
 \date{\today}

\begin{abstract}
The combination of fast propagation speeds and highly localized nature has hindered the direct observation of the evolution of shock waves at the molecular scale. To address this limitation, an experimental system is designed by tuning a one-dimensional magnetic lattice to evolve benign wave forms into shock waves at observable spatial and temporal scales, thus serving as a `magnifying glass' to illuminate shock processes. An accompanying analysis  confirms that the formation of  strong shocks is fully captured. The exhibited lack of a steady state induced by indefinite expansion of  a disordered transition zone points to the absence of local thermodynamic equilibrium, and resurfaces  lingering  questions on the validity of continuum assumptions in presence of strong shocks.
\end{abstract}

\maketitle

%-----------------

%\begin{document}

The propagation of shock waves in solids has received enormous attention in the last several decades \cite{mcqueen1960equation,holian1998plasticity,yao2019high,simmons2020quantum}. Experiments, molecular dynamic simulations, and continuum mechanics modeling, have been performed to investigate  shock waves \cite{catheline2003observation,espindola2017shear,chockalingam2020shear, Ramesh2008} and their interactions with complex material response such as plasticity \cite{chen2006dynamic}, damage  \cite{fensin2014dynamic}, dislocation and twinning \cite{higginbotham2013molecular,wehrenberg2017situ,turneaure2018twinning}, and  phase  transformation \cite{duvall1977phase,kadau2002microscopic,amadou2018coupling}. However, the microscopic mechanisms behind their formation are yet to be fully understood. 

Since the first development of modern shock wave theory, it is widely accepted that, at the continuum scale,  shock waves can be modeled as steadily propagating discontinuities within a medium \cite{salas2007curious}. While it is acknowledged that, in a physical system, even vanishing levels of viscosity or rate-sensitivity promote a continuous waveform, the thickness of this wave is thought to be steady and infinitesimal  compared to  the scale of the continuum process \cite{courant1948}. Hence, the main features of shock wave propagation can be captured using one-dimensional rate-independent theories \cite{von1950propagation}. However,  over the years, there have been indications of situations in which these assumptions breakdown \cite{hsu1974plastic, molinari2004fundamental, knowles2002impact, niemczura2011response,holian1978molecular}.  
Since the macroscopic response of a solid is intrinsically linked to its response at the microscopic scale, it is plausible that  in these situations additional information on the microscopic process occurring within the narrow region of the shock is needed to explain the continuum level observations. However, to the best of our knowledge, the evolution and propagation of shock waves at the molecular scale has only been captured via numerical simulations \cite{tsai1966shock,holian1978molecular,holian1979molecular}. Whereas their  direct observation can serve to better elucidate shock wave phenomena and to  distinguish between artifacts of numerical modeling and the actual physics.

Packed granular chains serve as an example  discrete system, which has been extensively studied due to its ability to   generate strongly nonlinear waves, including shock waves \cite{molinari2009stationary}, and Nesterenko solitary waves \cite{nesterenko1983propagation}. In these chains  the Hertzian  contact  between particles leads to their deformation in a highly nonlinear process, which is also responsible for  significant energy dissipation. The response of these systems is thus not directly comparable to molecular scale phenomena.

To mimic the molecular scale response,  we develop a desktop-scale  experimental realization of  shock wave evolution in a tunable magnetic lattice. We demonstrate the propagation of strong shocks and capture their entire evolution from a benign wave. Our validated numerical model provides  a comprehensive understanding of the observed phenomena and its sensitivity to both external damping and the imposed waveform. Moreover, it confirms  that this system supports the propagation of quasi-steady strong shocks, in which the shock front exhibits `soliton like' features propagating at constant velocity and strength, while the particle velocity profile reaches a steady oscillatory state. It is shown that for strong shocks a highly disordered transition regime emerges, from the shock front to the steady oscillatory state, and expands indefinitely. Thus, revealing  an unsteady feature of shock waves that nucleates at the molecular scale and can grow to the macroscopic scale.

%\textcolor{blue}{In sharp contrast to  packed granular chains, where the Hertzian  contact  between particles can induce strongly nonlinear waves, including shock waves \cite{molinari2009stationary} and Nesterenko solitary waves \cite{nesterenko1983propagation},  in the present system the highly nonlinear repelling forces between particles prohibit contact. This eliminates any effects of particle deformation, it thus results in  minimal energy dissipation and  permits direct comparison between theory and experiment. Additionally, large displacements of the particles enhances the observability of wave propagation phenomena in this system, and compares with predicted molecular scale response, as detailed next.}

 \begin{figure*}[ht!]
 \center
 \includegraphics[scale=0.45]{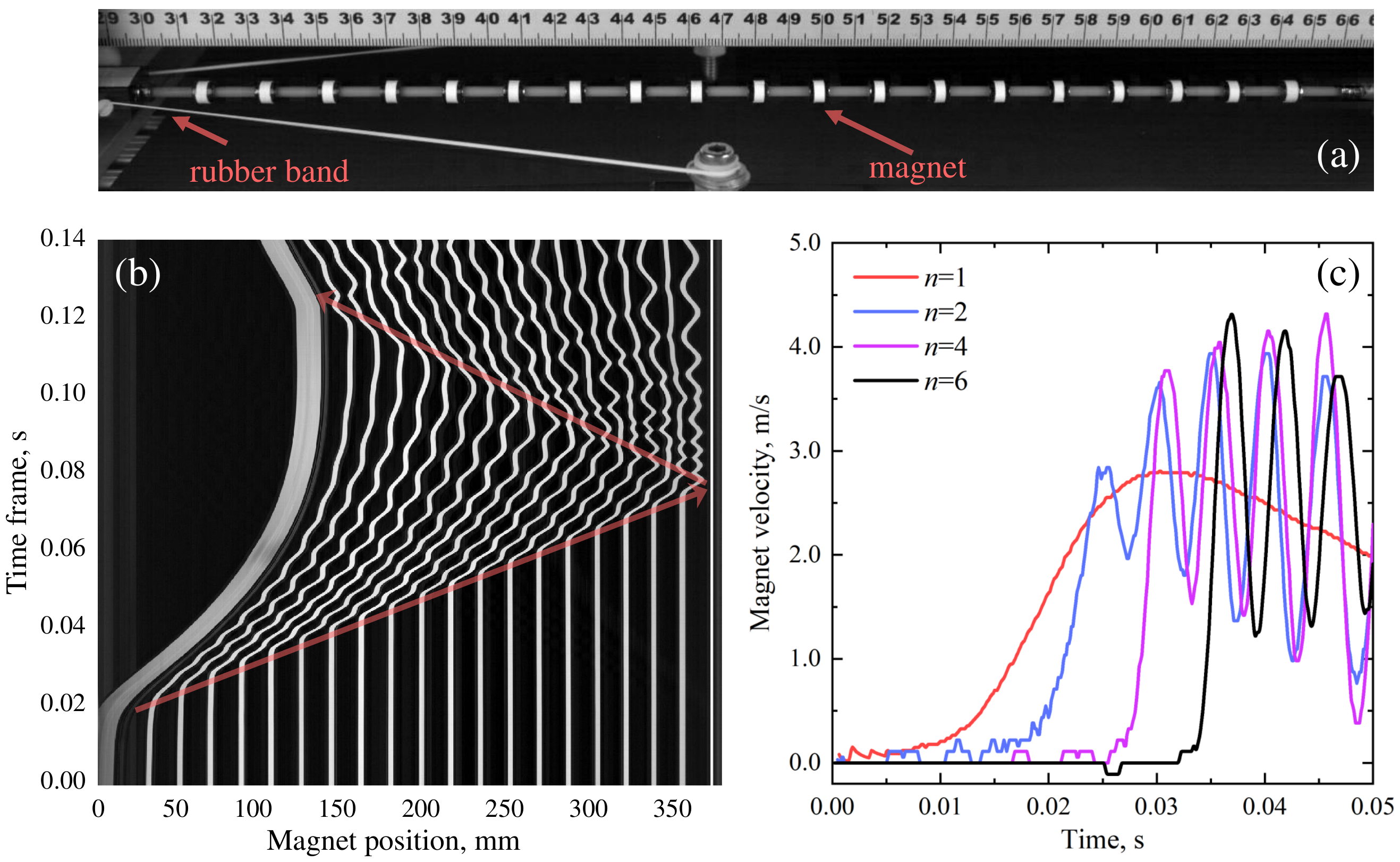}
 \vspace{-3mm}
 \caption{Experimental setup and representative results with initial magnet separation $r_0=12$  mm.
   (a) Experimental system consisting of  magnet lattice threaded on a rigid rod. The impact pulse wave is generated by releasing the left magnet. 
 (b) Magnet trajectories are shown as white curves, by vertically stacking images of the system at different times.
 (c) Magnet particle velocity profiles reconstructed via digital image correlation.
 }
 \label{fig.1}
 \end{figure*}
To realize shock wave evolution that is comparable to molecular-scale process, but  in a desktop-scale system, the experimental setup requires a tunable lattice with  minimal levels of dissipation.
Provided a finite imaging window, the system should  evolve a benign impact into a shock within a prescribed propagation distance, and at sufficiently slow velocities. 
The former can be achieved by  particles with highly nonlinear repelling forces (i.e. strongly convex force-separation curve), and the latter by tuning the  stiffness-to-mass ratio (i.e. the ratio between the local slope of the force-separation curve and the particle mass). To meet these requirements, we take advantage of the highly nonlinear repelling  nature of rare-earth magnets and  construct a lattice  of 21  particles with outer diameter of $6.35$ mm, length of $6.35$ mm, and mass of  $m=1.084$ g. As shown in Fig. \ref{fig.1}(a). The magnets are free to slide on a  non-magnetic, minimal friction, supporting cylindrical rod. The first magnet (on the left)  is attached to two tilted pre-stretched rubber bands through a plastic connector, while the last magnet is  fixed. The magnets are initially equi-spaced and pre-loaded to tune the repulsive force (or equivalently the stiffness) before impact. The imparted wave  is generated by releasing the first magnet, thus allowing the rubber-bands to contract and initiate the propagation (see Fig. 1(b) and \textcolor{blue}{\href{https://www.dropbox.com/s/2xkwkgsd73dxuqn/MovieS1.avi?dl=0}{Movie S1}}). The dynamic process is recorded by a high-speed camera (Photron SA5, 1024 x 1000 pixels) at 8 kHz, allowing the measurement of magnet displacement, velocity, and acceleration via digital image correlation method. Using this setup the impact strengths and lattice stiffness are separately tuned  by varying the pre-stretch  of the rubber bands, or the initial separation between the magnets, respectively. More details of the experimental system are given in Section S1 of the Supplementary Material.

To show that this system is capable of   evolving  an ordinary waveform into a shock, within the allocated propagation distance, we examine its response  to an impact.  The magnet trajectories  are shown in Fig. \ref{fig.1}(b)  for the case with maximum impact velocity $V_I=2.81$ m/s and with an initial magnet separation of $r_0=12$ mm. As indicated by the red arrow, the evolution of the magnet displacements shows  a wave propagating from the impacted end, into the lattice at  $V_P=6.42$ m/s. Then, upon arrival at the last magnet, a reflection wave propagates back.  It is seen from the displacement profiles that although the imparted wave form is  smooth; its propagation induces sharp oscillations in magnet particle displacement curves, indicating rapid changes in magnet velocities. 

If a shock forms, the wave profile is expected to steepen. By examining the velocity profiles of different  magnet particles in Fig. \ref{fig.1}(c), it is clearly observed that in our system   significant steepening occurs and is accompanied by oscillations that become more violent as propagation proceeds. In particular, notice the decreasing rise times (i.e. the duration from zero velocity to first peak velocity), which reduce from $23.9$ ms, for the first particle $(n=1)$, to $4.9$ ms, for $n=6$. This result clearly demonstrates the realization of a longitudinal shock wave and its evolution from a simple wave. Moreover, the  violent oscillations of increasing amplitude, in what seems to be a highly disordered process, are indicative of strong shocks.

\begin{figure}[ht!]
 %\vspace{-1mm}
 \center
 \includegraphics[scale=0.3]{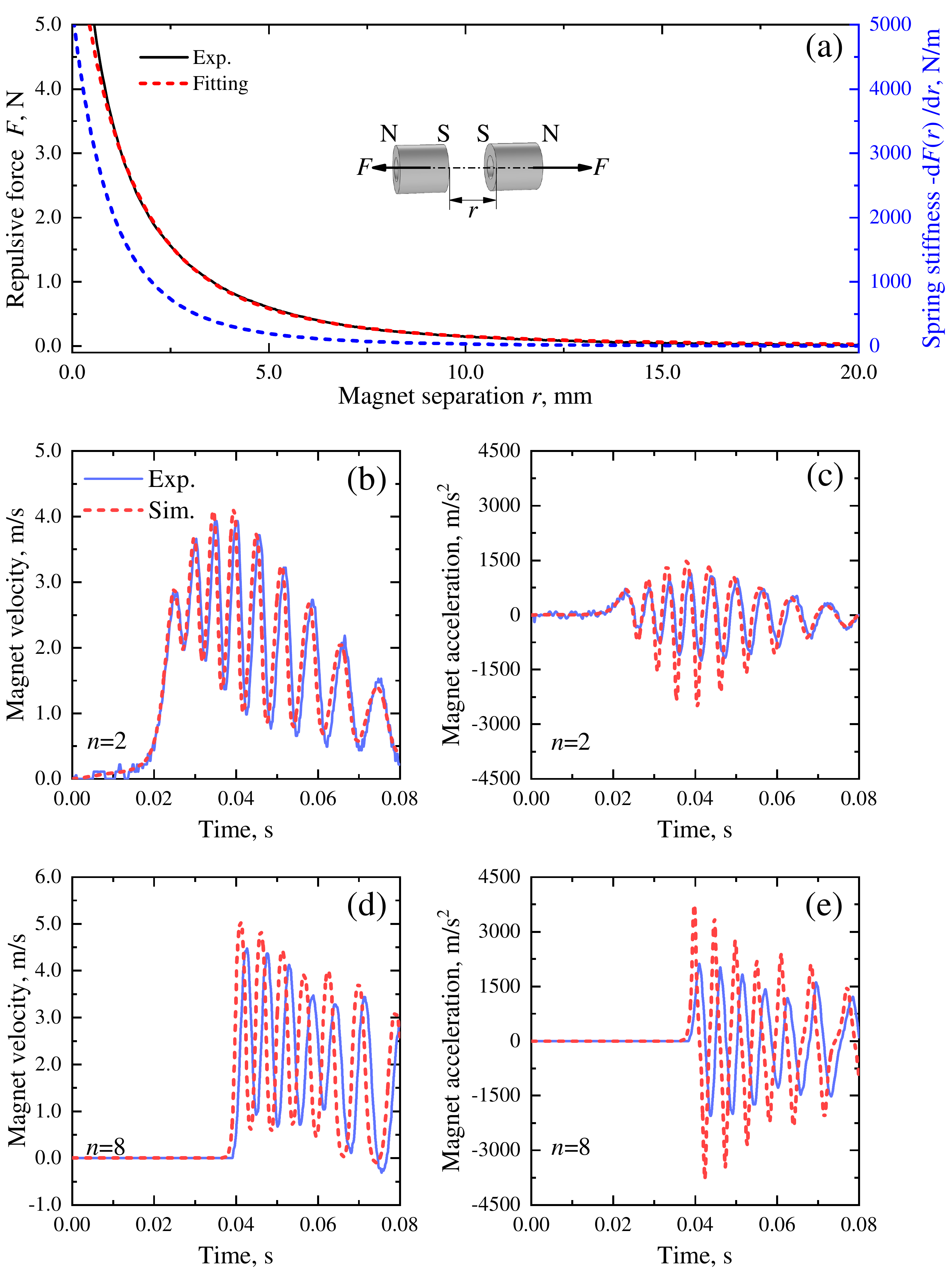}
 \caption{Comparison of experimental and numerical results. (a) Force-displacement curve and instantaneous stiffness for the nearest magnet interaction. (c-d) Experimental (continuous blue lines) and numerical (dashed red lines) results for magnet particle velocities and accelerations.}
 \label{fig.2}
 \end{figure}

To better understand the observed shock evolution, we numerically model the system as a chain of particles connected by nonlinear springs.  %Since the ring magnet is much stiffer than the magnetic nonlinear spring, it is assumed to be a rigid particle, and its 
%The length of the magnet is not considered in the calculation of propagation velocity in following discussions.
In the following analysis, we only consider the interaction between first neighboring magnets, and the length of the magnet is not considered in the calculation of propagation velocity. Although some influence may arise from the magnetic field of the non-nearest particles, it is a second order effect (see Section S2 in the Supplementary Material).
Accordingly,  the equation of motion for the $n^{\text{th}}$ magnet reads  
\begin{equation}\label{eq.1}
m\frac{{\rm d}^2 u_{n}}{{\rm d}t^2}=F_{n-1}-F_{n}-f_{n},
\end{equation}
where $u_n$ denotes the particle  displacement, and  $F_n$, $f_n$ are the repulsive and frictional forces, respectively. In particular, based on experimental measurement of the force-dispacement curve (Fig. \ref{fig.2}a), the repulsive force  is approximated using the formula $F=K/(r+b)^q$, where $r$ is the separation between two neighbor particles, and the coefficient values are ${K}=413.8$ Nmm\textsuperscript{3}, $b=3.917$ mm,   $q=3$. A Coulomb model captures the influence of friction between the rod and the magnets via the formula $f_n=\mu(mg+p(F_{n-1}+F_n))$, where the coefficients $\mu=0.285$ and $p=0.012$ are experimentally measured    (see Supplementary Material S3), and  $g$ is the  gravitational acceleration. The  measured motion of the impacting magnet, $u_1(t)$,  is given as a boundary condition at one end, while at the other end we impose ${u}_{21}(t)=0$. Numerical results obtained using this model are compared with  experimental curves for the $2^{\text{nd}}$ and $8^{\text{th}}$ magnets  in Fig. \ref{fig.2}(b-d), and show excellent agreement for  the velocity profiles. The acceleration profiles are also well captured by the simulation. The discrepancy in peak accelerations can be explained by the limited image resolution ($\sim20$ pixels per ring magnet particle length), which is insufficient to capture sharp changes in acceleration. 

\begin{figure}[ht!]
 \center
 \includegraphics[scale=0.3]{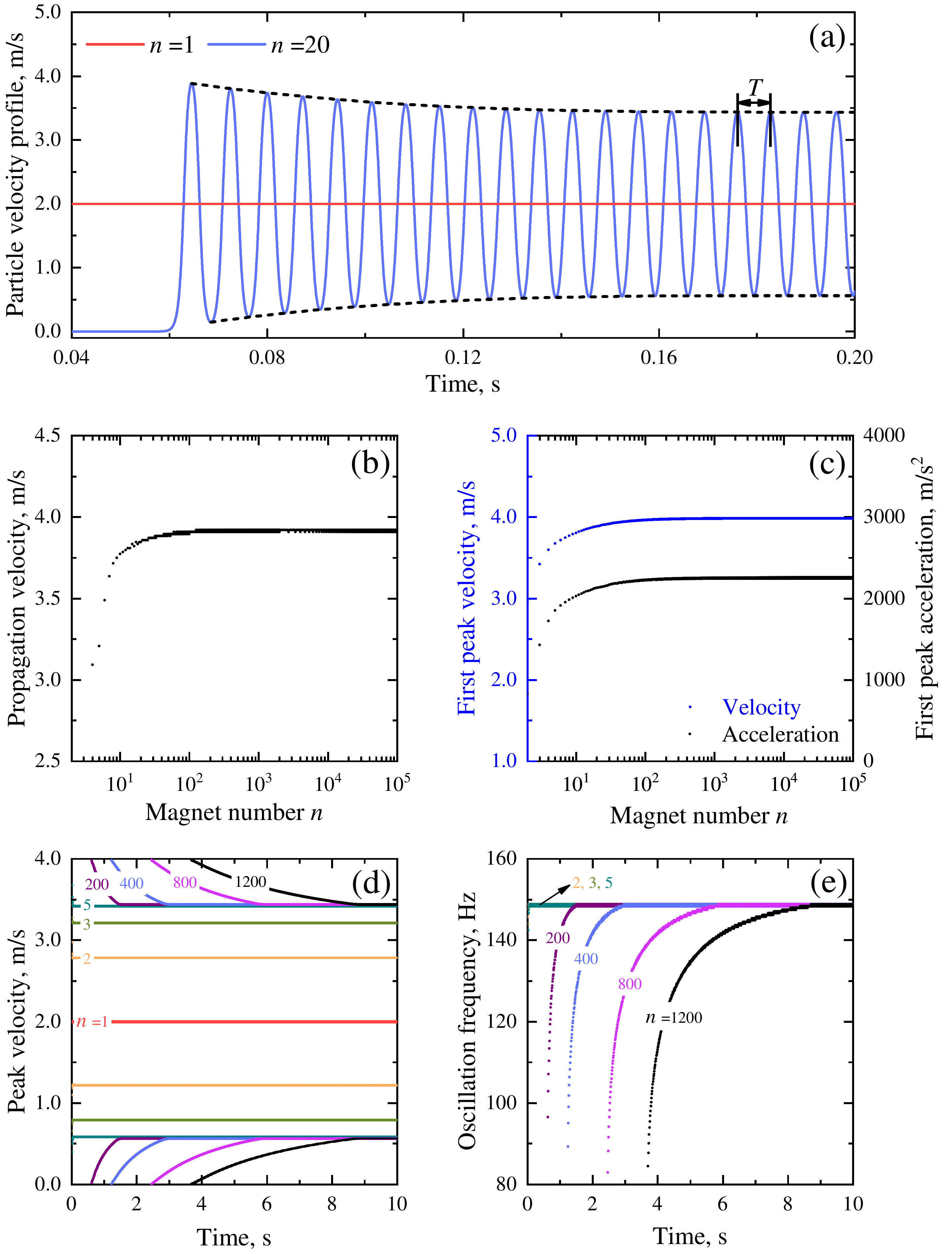}
 \caption{Long-time dynamic response of the system without friction. (a) Typical velocity profile of a particle. (b) Variation of propagation velocity as the wave progresses through particles into the lattice. (c) First peak velocity and first peak acceleration of each particle. (d) Evolution of peak velocity (both local maximum and minimum peaks) with time for various particles. (e)  Evolution of oscillation frequency  with time for various particles. Results are shown for the system with initial separation $r_0=12$ mm.}
 \label{fig.3}
 \end{figure}
Next, we use our calibrated model to investigate the long-time behavior of strong shocks. For simplicity, we consider a long  lattice subjected to a constant impactor velocity (a long  lattice is used to avoid wave reflections). Fig. \ref{fig.3}(a) shows a typical velocity profile obtained for an impactor velocity of $V_I=2$ m/s, in absence of friction. Upon arrival of the shock front, the particle velocity is shown to rapidly increase to $3.87$ m/s, followed by  strong oscillations with a decaying amplitude. Unlike a linear system,  for which the vibration amplitude decays completely (see Supplementary Material S4 and S7), stable finite-amplitude oscillation about the impactor velocity is eventually attained.  Note that this  motion is  non-harmonic  due to the nonlinearity of the system. Examining the corresponding propagation velocity of the shock front in  Fig. \ref{fig.3}(b) we show that it gradually approaches a constant value of $V_P=3.92$ m/s. It is notable that this stabilized propagation velocity is significantly larger than the impactor velocity   and the linear propagation velocity $V_0=1.62$ m/s \footnote{Here the linear velocity refers to the  propagation velocity of small amplitude disturbances, which can be calculated as ${V_0}={r_0}\sqrt{{-(1/m){\rm d}F(r)}/{{\rm d}r}|_{r=r_0}}$, where $r_0 $ represents the initial magnet separation.}. If frictional effects are included, a gradual decay of the propagation velocity is expected beyond a peak value (see Section S5 in the Supplementary Material). Nonetheless,  once  developed, the early time propagation velocity (i.e. for the first $\sim20$ magnets) is comparable to the constant propagation velocity in the frictionless system.
Fig. \ref{fig.3}(c) shows that both the first peak velocity and the first peak acceleration increase with  increasing particle number, which is an intrinsic feature of strong shock waves. Eventually, the competition between  nonlinearity and dispersion in the system results in  saturation of the  first peak velocity and peak acceleration. In particular, the saturated first peak velocity is 2$V_I$.

Further, to understand the transition from the wave front to the stabilized oscillatory state, Figs. \ref{fig.3}(d,e) present the peak velocity and the corresponding oscillation frequency for different particle numbers as a function of time. We find that both the amplitude and the period decrease with time; moreover, after a rapid increase in stabilized oscillation amplitude (from the $1^{\text{st}}$ magnet to the $5^{\text{th}}$ magnet), the following particles arrive at the same oscillatory state (same amplitude and same period). Nonetheless it is observed that the time of transitioning from peak velocity to the stabilized state is longer for the larger particle numbers, resulting in the highly disordered transition zone that expands indefinitely as the shock front penetrates deeper into the undisturbed lattice (see \textcolor{blue}{\href{https://www.dropbox.com/s/f0gkck8ulcsfh37/MovieS2.avi?dl=0}{Movie S2}}). Analogous to the interpretation of molecular scale response, the finite amplitude steady oscillation in the wake of a shock is consistent with an increase in  temperature \cite{straub1979molecular}, whereas the disordered transition region appears to be out of thermodynamic equilibrium and its growth can be attributed to increasing entropy.

\begin{figure}[ht]
 \center
 \includegraphics[scale=0.32]{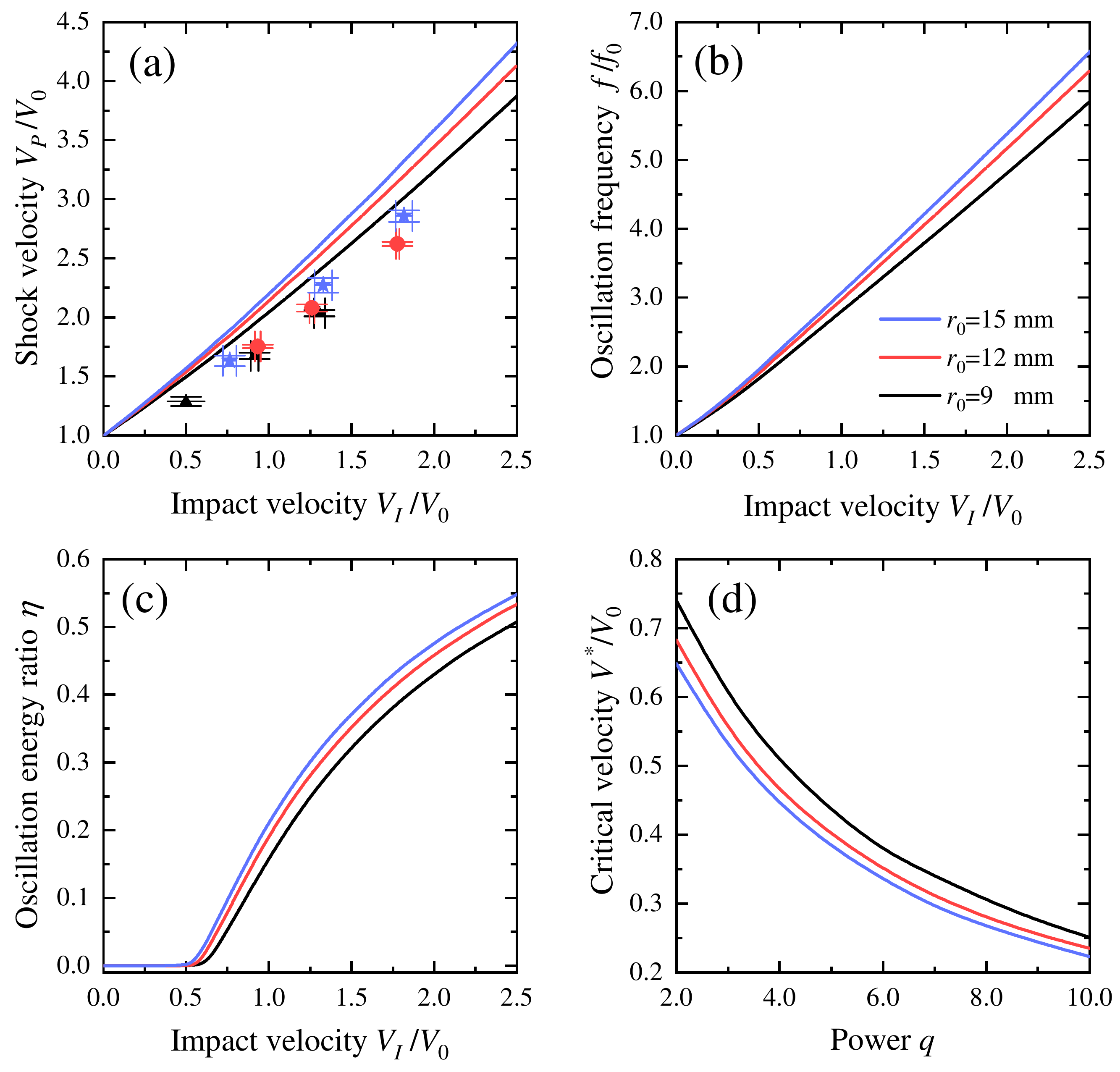}
 \caption{Dependence  of the stabilized shock response on impact velocity. Curves represent numerical solutions. Experimental datum are shown as colored markers  with error bars.
 (a) Shock wave propagation velocity. (b) Stabilized oscillation frequency, $f$. (c) Oscillation energy ratio, $\eta$.  (d) Critical velocity. Here $f_0$ represents the linear  oscillation frequency, defined as
 ${f_0}=1/\pi\sqrt{{-(1/m){\rm d}F(r)}/{{\rm d}r}|_{r=r_0}}$, and $\eta=1/T\int_{t_0}^{t_0+T} (V(t)/V_I-1)^2 \,{\rm d}t$, where  $T=1/f$. If the oscillation is harmonic,  $\eta=0.5$.
}
 \label{fig.4}
 \end{figure}

To further understand the range of shock wave response realized in our experiments, we explore the effect of the impactor velocity on the propagation velocity of the  quasi-steady shock wave in Fig. \ref{fig.4}(a). The nearly linear dependence   observed in both experimental and numerical results   resembles the reported experimental findings of shock Hugoniot data in metallic materials \cite{marsh1980lasl} and molecular simulations of shock waves \cite{hill1980propagation}.  The agreement between theory and experiments is shown with slight deviations attributed primarily to effects of   friction, and the precise form of the imparted wave that are neglected in the simulation (see Supplementary Material S5, S6). While these curves,  as well as the corresponding oscillation frequency (Fig. \ref{fig.4}b), do not reveal information on the shock strength, we examine in Fig. \ref{fig.4}(c) the  kinetic energy associated with the steady state oscillation. Quite noticeably, the increase in oscillation energy becomes pronounced  beyond a critical impactor velocity, $V^*$,  which is smaller than the linear propagation velocity (i.e. $V^*<V_0$). This  threshold velocity represents the transition into the strong shock regime, which is characterized by  increased  energetic consumption. With $V^{*}/{V_0}\sim 0.5$ in our experimental system, it is clear that strong shocks are observed (see Fig.\ref{fig.4}(a)). %Note that the oscillation  energy ratio can be larger than 0.5, indicating the nonharmonicity of the local oscillation.   
In Fig. 4(d) we further investigate this critical threshold by examining  the influence of the stiffening law, or in particular, the power $q$. We observe that a system with increased stiffening can be driven beyond the critical threshold by lower impactor velocities.

In conclusion, we have shown that the desktop-scale experimental system presented here allows for  complete spatio-temporal capture of the evolution of strong shocks from benign imparted wave forms. This is facilitated by taking advantage of the highly nonlinear repelling force between  neighboring rare-earth magnets  in a tuneable one-dimensional lattice. Comprehensive investigation of the lattice response uncovers behaviors of strong shocks, that agree with predictions from Molecular Dynamic (MD) simulations. Hence,  this work gives rise to a new avenue for investigation of shock wave phenomena at the microscopic scale. Moreover, through this analysis, we observe the formation of a highly disordered  transition region  in the wake of strong shocks.   This region nucleates at the particle scale, but continues to grow indefinitely. Observation of this phenomena at the macro-scale, raises questions on the validity of continuum assumptions in the presence of strong shocks. Future work can take advantage of this system to expand beyond uniaxial propagation and can include additional physical effects, such as structure defects, thermal vibrations, and dissipation. Moreover, it is worth mentioning that the current design could also be modified to explore other nonlinear wave phenomena, such as solitons \cite{deng2017elastic,deng2019focusing,zhang2019programmable}, elastic bandgaps \cite{deng2018metamaterials}, and nonreciprocal waves \cite{wang2018observation,nassar2020nonreciprocity,brandenbourger2019non}.

\bigskip
The authors wish to acknowledge the financial support of the Army Research Office, under award number W911NF-19-1-0275, and support from the National Science Foundation (CMMI, MOMS, 1942016).

\bibliography{bibliography.bib}

\clearpage

\onecolumngrid

\renewcommand{\thepage}{S\arabic{page}} 
\renewcommand{\theequation}{S\arabic{equation}} 
\renewcommand{\thesection}{S\arabic{section}}  
\renewcommand{\thetable}{S\arabic{table}}  
\renewcommand{\thefigure}{S\arabic{figure}}

\setcounter{page}{1}
\setcounter{figure}{0} 
\setcounter{equation}{0}
%\begin{document}

\begin{center}
\Large Observation of ultra-slow shock waves in a tunable magnetic lattice

Supplementary Material
\end{center}

\noindent\makebox[\linewidth]{\rule{0.8\paperwidth}{0.4pt}}
%\\ Evolution of ultra-slow shock waves observed in a tunable magnetic lattice}

%\author{Jian Li}
%\affiliation{ Massachusetts Institute of Technology, Department of Civil %and Environmental Engineering, Cambridge, MA, 02139, USA}
%\author{Senthilnathan Chockalingam}
%\affiliation{ Department of Aeronautics and Astronautics, Massachusetts %Institute of Technology, Cambridge, MA 02139, USA}
%\author{Tal Cohen}
%\email{Corresponding author: talco@mit.edu}
%\affiliation{ Massachusetts Institute of Technology, Department of Civil %and Environmental Engineering, Cambridge, MA, 02139, USA}
%\affiliation{ Massachusetts Institute of Technology, Department of %Mechanical Engineering, Cambridge, MA, 02139, USA}

\section{Experimental details}

\begin{figure}[ht!]
 \center
 \includegraphics[scale=0.65]{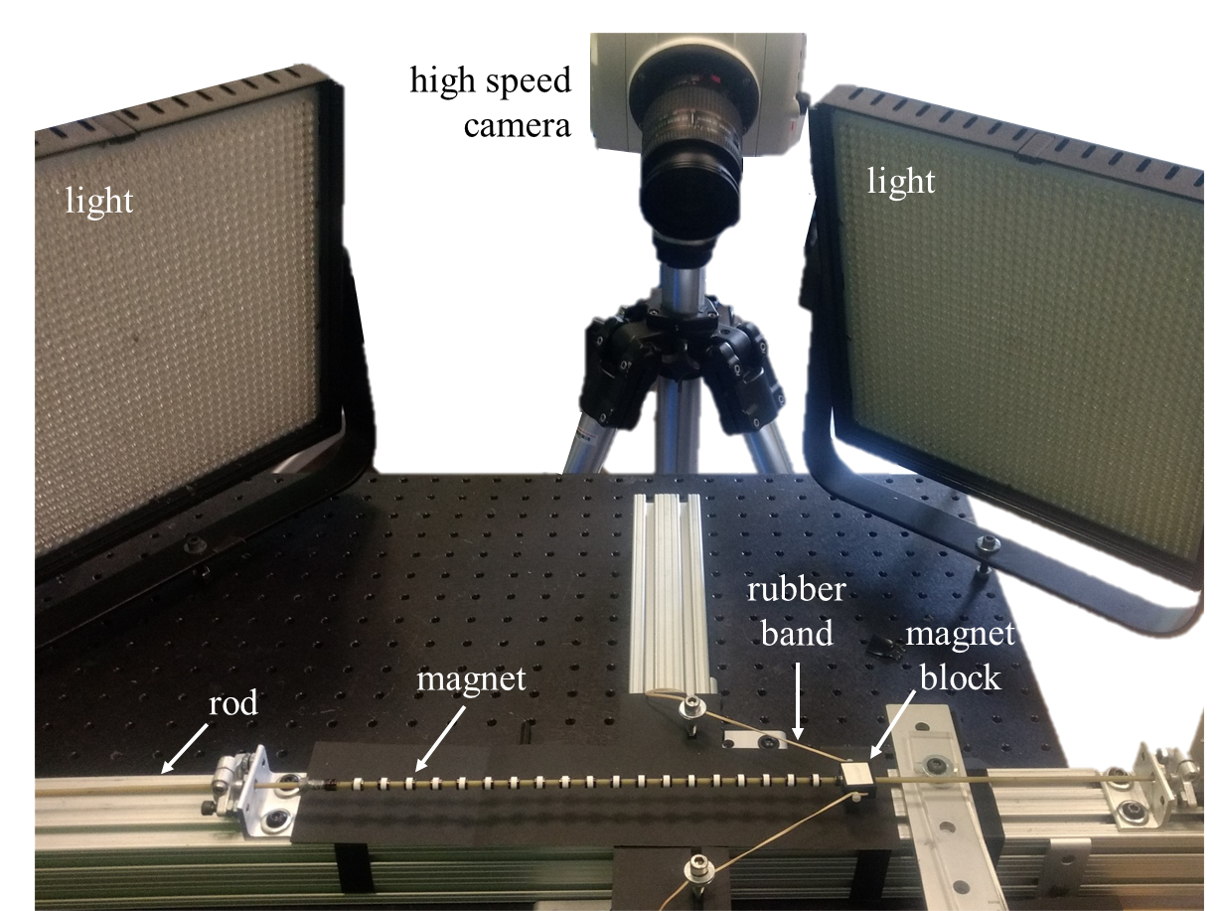}
  \caption{
 Experimental setup.
 }
 \label{fig.s1}
 \end{figure}

The experimental setup is shown in Fig. \ref{fig.s1}. The system consists of a lattice of ring magnet particles, which is threaded along the rod such that the polarity of neighboring magnets has an  opposite orientation (i.e.  [N-S]-[S-N]-...-[N-S]-[S-N]), thus exerting a repulsive force. Each magnet particle is composed of two ring magnets with outer diameter of 6.35 mm, inner diameter of 3.175 mm, length of 3.175 mm, and mass of 0.542 g (KJ Magnetics Inc, Grade N42, R422). The rod has an outer diameter of 3.150 mm and is made of epoxy resin with fiberglass fabric reinforcement (McMaster-Carr, High-Temperature Garolite G-11 Rod). The rod with fixed ends on an aluminum alloy rail, is pre-stretched to minimize bending deformations. The last magnet on the left side is fully fixed to the rod. The first magnet on the right side is attached to two tilted pre-stretched rubber bands through a plastic connector. The  magnets are disturbed to be equispaced. In addition, each magnet is covered by a  white label, to enhance contrast for optical tracking. Finally, a pulse wave is generated by releasing the first magnet, which then propagates into the lattice. The  dynamic process is recorded using a high-speed camera (Photron SA5, 1024 x 1000 pixels) at 8 kHz.
 
 \begin{figure}[ht!]
 \center
 \includegraphics[scale=0.55]{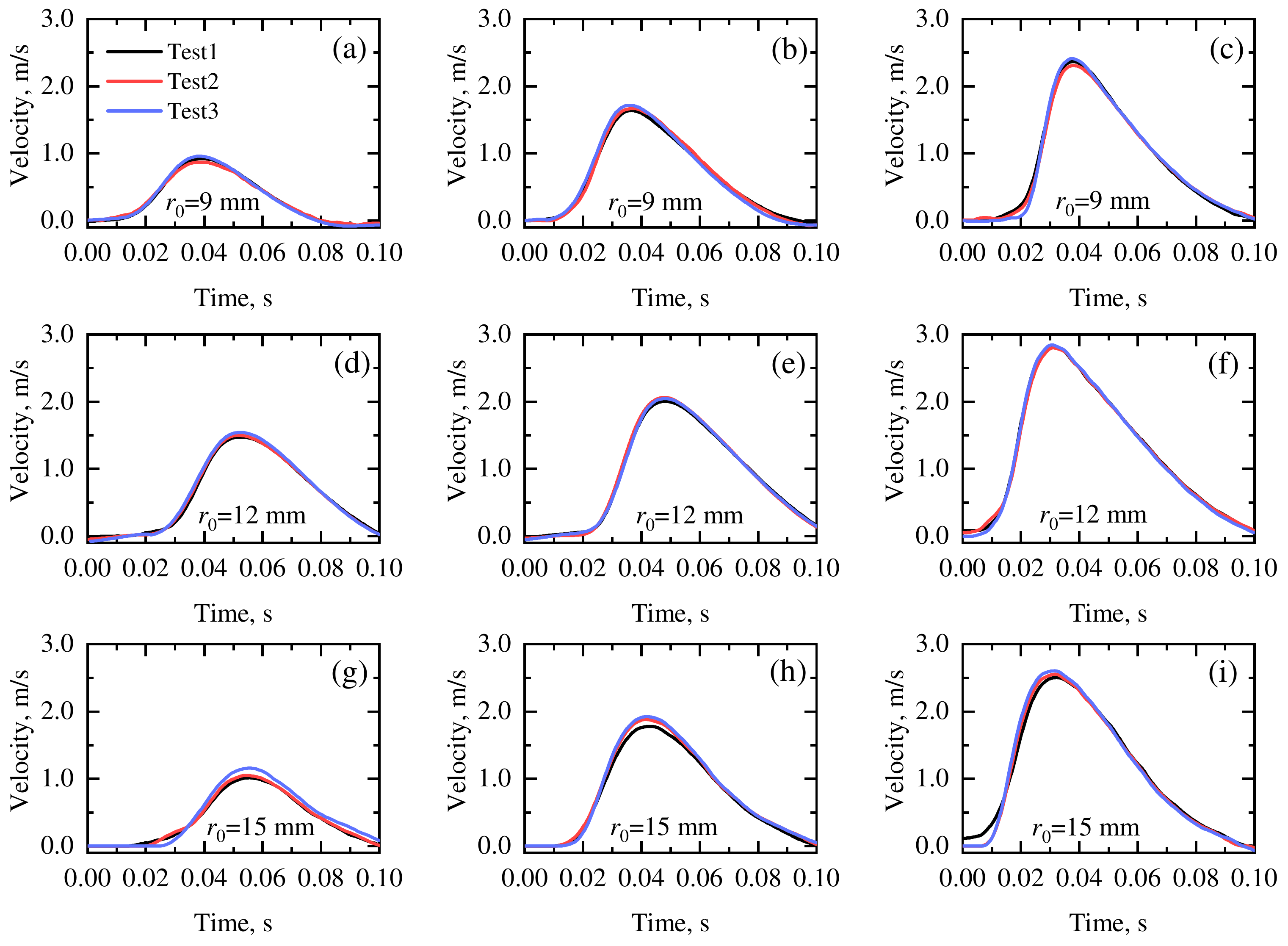}
  \caption{
 Velocity profiles of the impact signal. Different rows show profiles obtained with different initial separations and thus represent systems with different initial stiffnesses. 
 Different columns show profiles obtained for different impact strengths.
 }
 \label{fig.s2}
 \end{figure}
In experiments, different impact strengths are tuned by adjusting the pre-stretch  of the rubber bands, and thus the stiffness of the system (by modifying the separation between magnets). Note that each set of experiment is repeated three times. The impactor velocity signals (i.e. the velocity profiles of the $1^{\text{st}}$ magnet) are shown in Fig. \ref{fig.s2}.

 \clearpage

 \section{Measurement of the magnetic interaction force}
\begin{figure}[ht!]
 \center
 \includegraphics[scale=0.5]{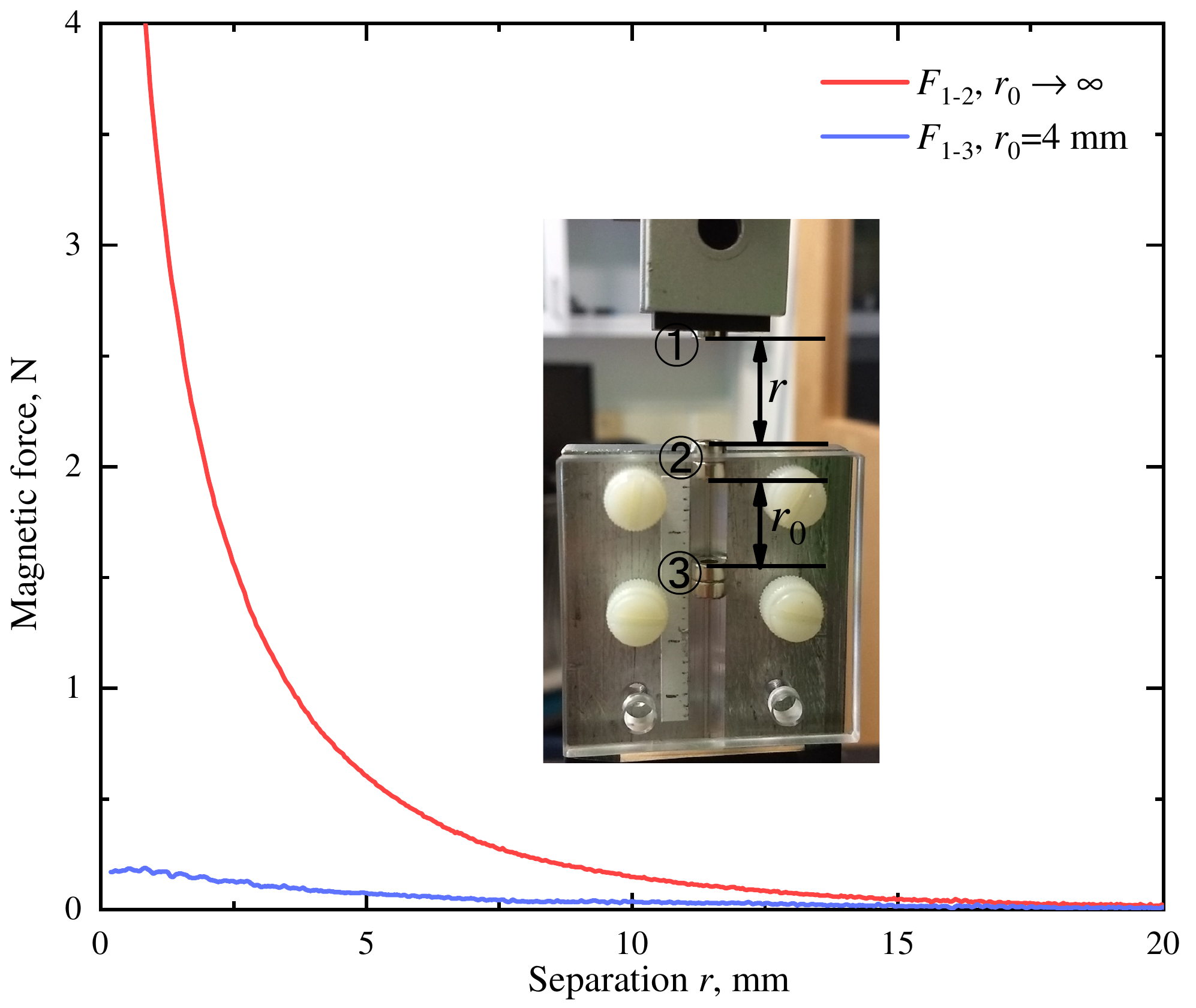}
  \caption{
  Magnetic force for the nearest interaction $F_{1-2}$ and the 2th nearest interaction $F_{1-3}$.
  }
 \label{fig.s3}
 \end{figure}

 To quantify the magnetic interaction force, we develop a setup to measure the $1^{\text{st}}$ neighbor and $2^{\text{nd}}$ neighbor interactions. As shown in Fig. \ref{fig.s3}, three magnets are aligned with opposite polarity (i.e. [N-S]-[S-N]-[N-S]) as in the chain magnet, thus the nearest interaction is a repulsive force, while the $2^{\text{nd}}$ nearest interaction is an attractive force. The gap between the $2^{\text{nd}}$ magnet and the $3^{\text{rd}}$ magnet is fixed at a prescribed separation $r_0$. Using an  Instron universal testing instrument we displace the  $1^{\text{st}}$ magnet along the vertical direction and measure the interaction force. We then employ  the principle of superposition to recover the magnet force for the nearest interaction (i.e. with $r_0\to \infty$) and $2^{\text{nd}}$ nearest interaction.  Fig. \ref{fig.s3} shows that the $2^{\text{nd}}$ nearest interaction force is about 5\% of the nearest interaction force, thus the non-nearest interaction is not considered in the numerical model.

 \section{Calibration of the Coulomb friction coefficient}
\begin{figure}[ht!]
 \center
 \includegraphics[scale=0.32]{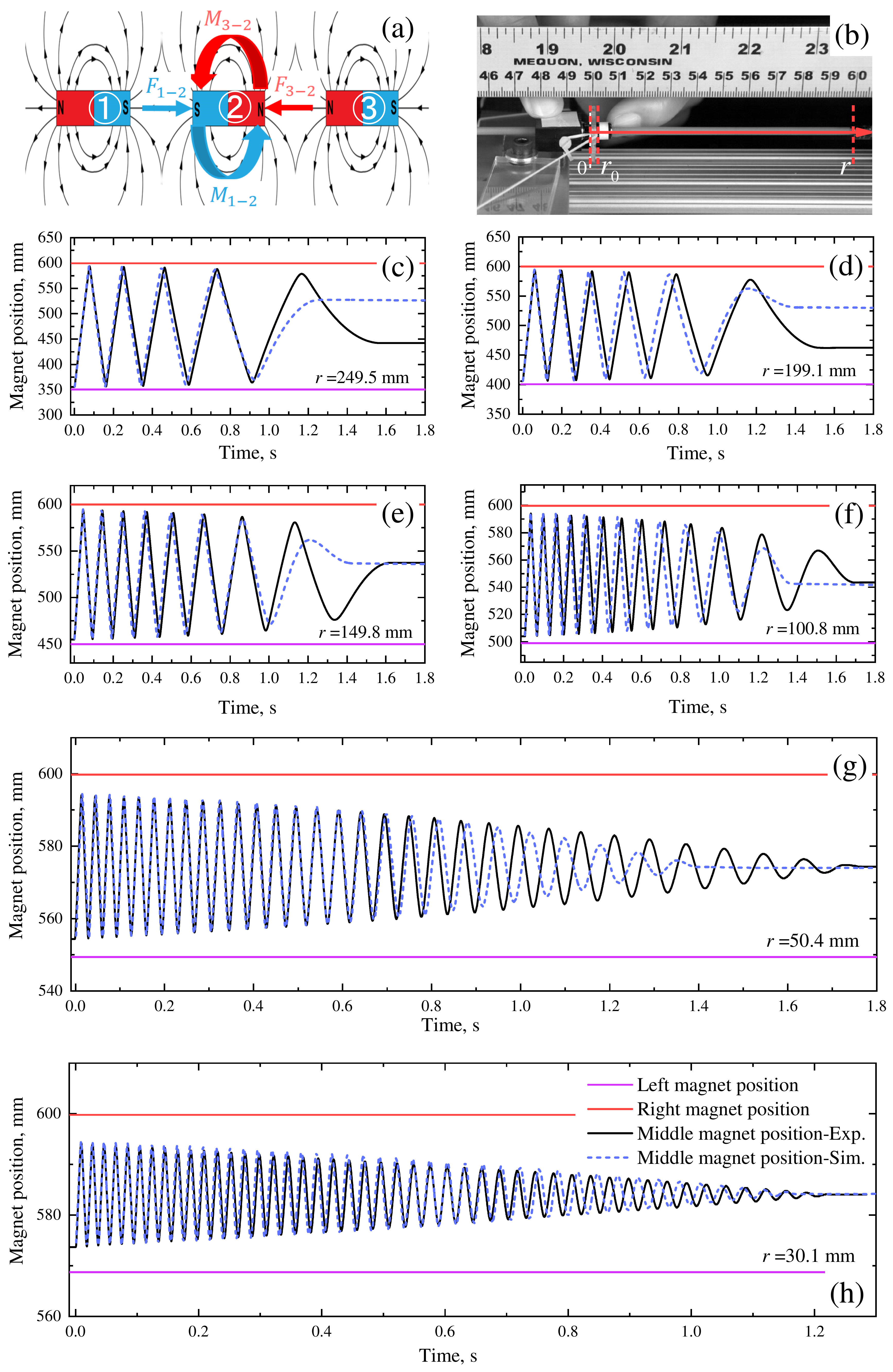}
  \caption{
Calibration of the Coulomb friction coefficient. (a) Schematic illustration of the exerted force and moment by the nearest magnet. (b) Calibration experiment setup. (c-h) Comparison of experimental (continuous back curves) and numerical (dashed blue curves) results on the central magnet position under different initial magnet separations. The numerical results are obtained with friction coefficients $\mu=0.285$ and $p$=0.012. }
 \label{fig.s4}
 \end{figure}

 Here we quantify the value of friction in our experimental system. The Coulomb friction model, which assumes that the friction force is proportional to the normal contact force, is employed. Fig. \ref{fig.s4}(a) illustrates the magnetic forces acting on a particle within the lattice. Since the non-nearest interaction is much weaker than the nearest interaction (as shown in Fig. \ref{fig.s2}), only the nearest interaction force is considered. Hence, along the horizontal direction, the repelling forces are $F_{1-2}$ and $F_{3-2}$. Additionally, the repelling/attracting force couples induce  a torque  which is  denoted by $M_{1-2}$ and $M_{3-2}$, respectively.  Given the horizontal orientation of the lattice, gravitation induces a normal force  $mg$, where $m$ is the particle mass. Reactions from the suspending rod constrain the particle against rotation and translation in the vertical direction. The contact between the particle and the rod adds to the normal force,  which contributes to the frictional force and thus  resists to  sliding of the magnets. We model the normal force as an additive contribution of the gravitational force and the reaction to the torque, which is proportional to the sum of the repelling forces (i.e. ${F}_{1-2}+{F}_{3-2})$ as 
\begin{equation}\label{eq.1_diff}
{f}_2=\mu(mg+p({F}_{1-2}+{F}_{3-2})),
\end{equation} 
 where  $\mu$ is the Coulomb friction coefficient and $p$ is an unknown dimensionless scaling coefficient. Both are determined via a dedicated  calibration method, as detailed next.  First, as shown in Fig. \ref{fig.s4}(b), we conduct an experiment in the system with three magnet particles. In particular,  the left magnet and the right magnet are fixed at a certain separation $r_1$, the central magnet is displaced towards the left magnet with a separation $r_0$. Then, upon removing the applied force, the central magnet oscillates between two fixed magnets with a decaying amplitude (due to the existence of friction) until reaching a static equilibrium. In calibration experiments, $r_0$ is set to be constant (i.e. $r_0=1.8$ mm), while $r_1$ varies from $249.5$ mm to $30.1$ mm.
 Finally, using the hypothesised friction law  (i.e. Eq. \eqref{eq.1_diff}) we solve the equation of motion for the one-degree-of freedom damped vibration of the central magnet, using the appropriate boundary conditions. The coefficients are then estimated by fitting to the experimentally measured results. Through this procedure, we obtain  $\mu=0.285$ and $p=0.012$. Fig. \ref{fig.s4}(c-h) shows the comparison between experimental and numerical results with the fitted coefficients. We observe an excellent agreement over a wide range of oscillation amplitudes.

 \clearpage

 \section{Dynamic response of a linear lattice}
\begin{figure}[ht!]
 \center
 \includegraphics[scale=0.4]{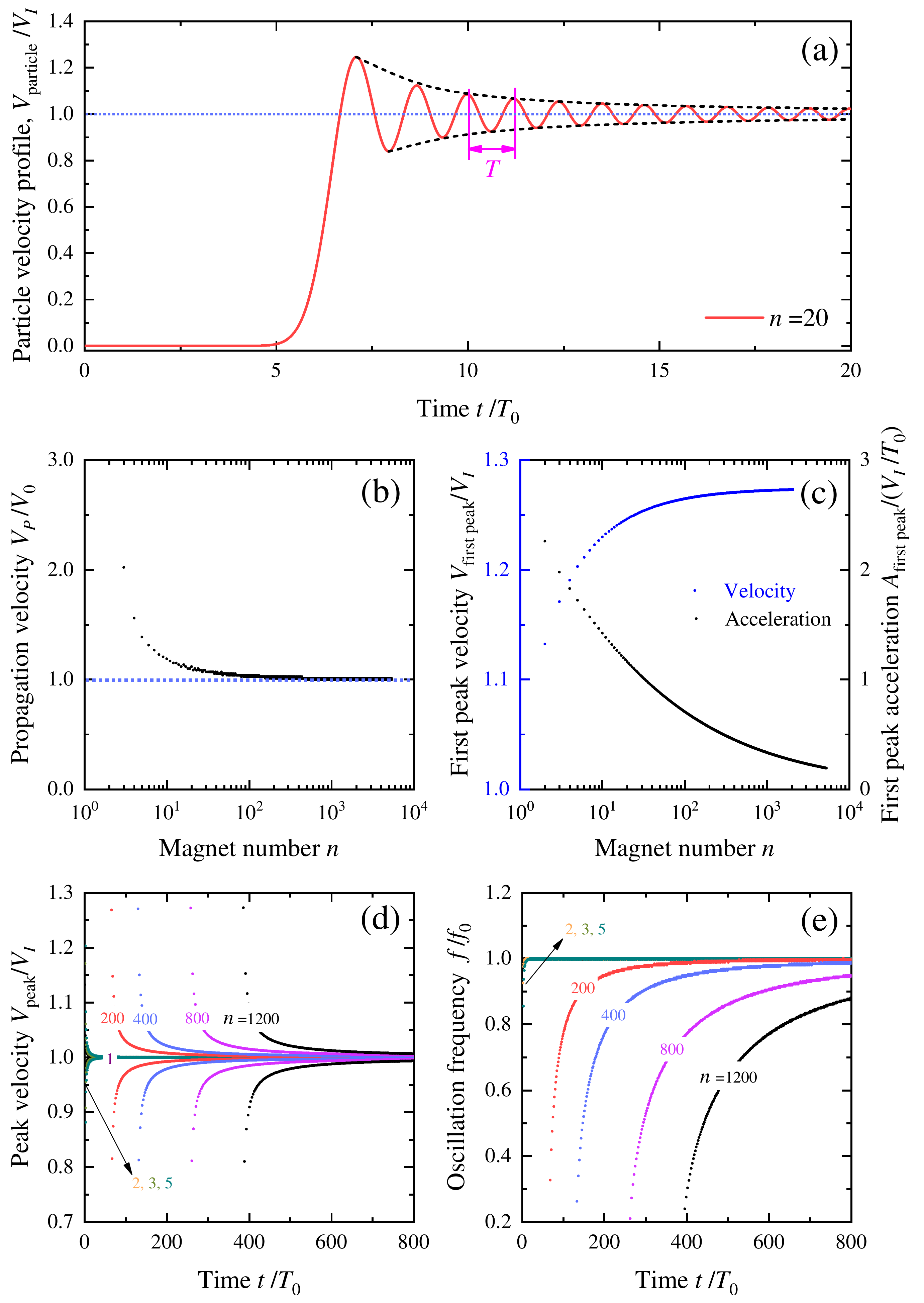}
  \caption{ Dynamic response of the linear lattice. (a) Typical velocity profile for a linear chain system under constant impact velocity. Dependence  of  propagation  velocity  (b),  first  peak  velocity  and  first  peak acceleration (c) on particle number.  Evolution of peak velocity (d) and oscillation frequency (e) with time for various particles. Note that for the linear chain system with stiffness $k$ and particle separation $r_0$, the linear propagation velocity $V_0=r_0\sqrt{k/m}$, oscillation frequency $f_0=1/\pi\sqrt{k/m}$, oscillation period \emph{T}\textsubscript{0}=1/\emph{f}\textsubscript{0}.}
 \label{fig.s5}
 \end{figure}

 For comparison with the nonlinear response provided in the main text, in  Fig. \ref{fig.s5} we provide results for  a linear lattice under constant impact velocity. Here, the linear lattice implies that the interaction force between neighboring particles is proportional to the particle separation, namely $F=kr$, and $k$ represents the stiffness.  As shown in  Fig. \ref{fig.s5}(a)  for $n=20$, upon arrival of the wave front,  the particle velocity rapidly increases to a peak value, then oscillates with a decaying amplitude about the impactor velocity. Moreover, the propagation velocity gradually decreases and asymptotically approaches  the analytical linear wave velocity (see Fig. \ref{fig.s5}(b)). In terms of first peak velocity and acceleration, Fig. \ref{fig.s5}(c) shows that the larger magnet number has a larger first peak velocity, but a smaller first peak acceleration, which is opposite to nonlinear stiffening system (the larger magnet number has a larger acceleration, see Fig. 3c). In Fig. \ref{fig.s5}(d) and (e), we find that both the oscillation amplitude and the period decrease with the increase of propagation time; in particular, the oscillation amplitude eventually decays completely.

  \section{Effect of friction on propagation velocity}
\begin{figure}[ht!]
 \center
 \includegraphics[scale=0.45]{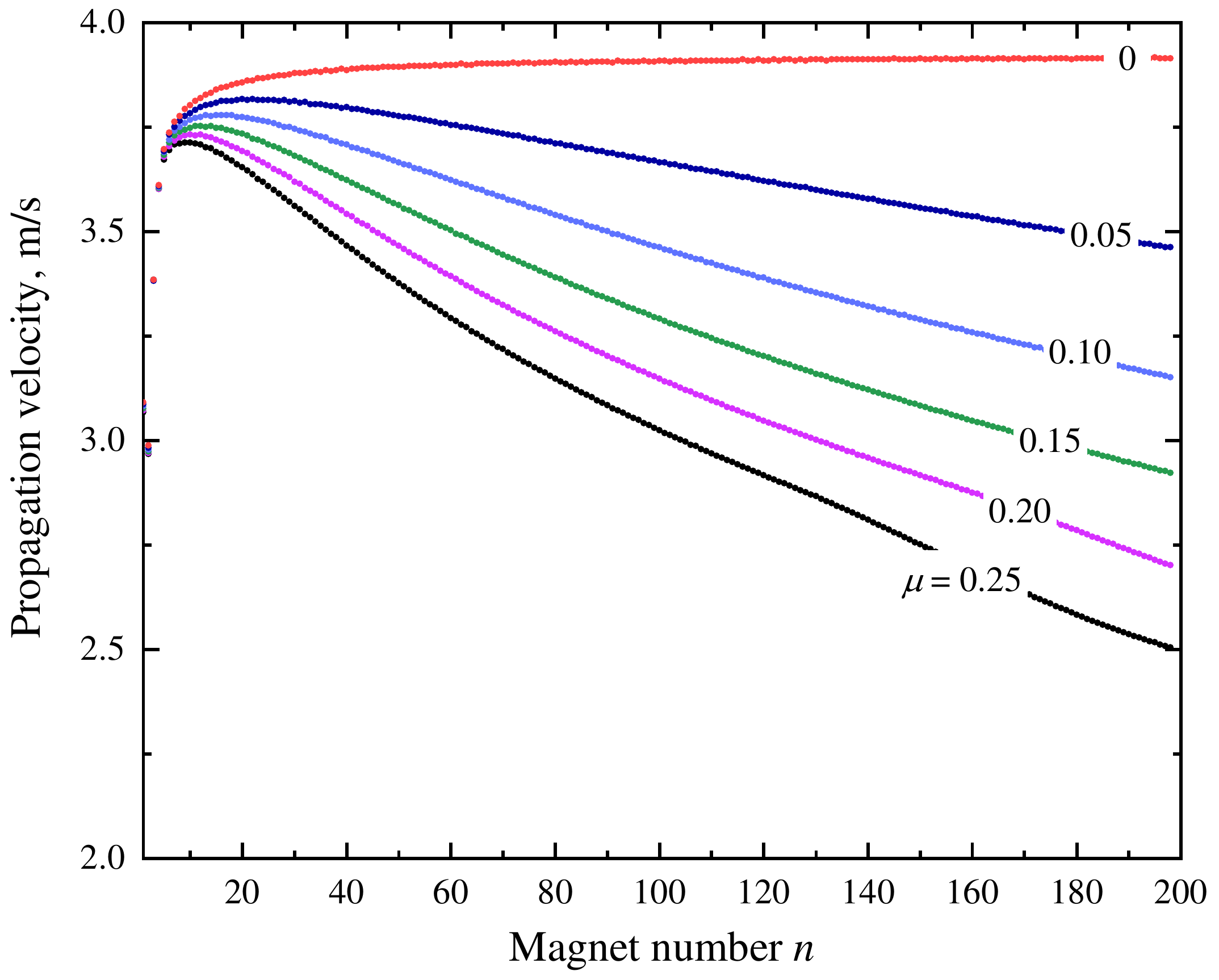}
  \caption{
Effect of friction on propagation velocity. Results are obtained by the calibrated simulation model with the approximated friction force law (i.e. Eq.(\ref{eq.1_diff})) and initial magnet separation $r_0=12$ mm under constant impact velocity $V_I=2$ m/s.}
 \label{fig.s6}
 \end{figure}

  Fig. \ref{fig.s6} reports the effect of friction on the propagation velocity in the nonlinear lattice. Our results reveal that the friction dramatically affects the propagation velocity. When no friction is considered, the propagation velocity eventually approaches a constant value. However, even vanishing levels of friction eventually lead to decay of the propagation velocity. In addition,  systems with larger values  of the friction coefficient exhibit a more dramatic decrease in the propagation velocity.

 \vspace{1mm}
 \section{Effect of loading profile in the nonlinear system}
\begin{figure}[ht!]
 \center
 \includegraphics[scale=0.36]{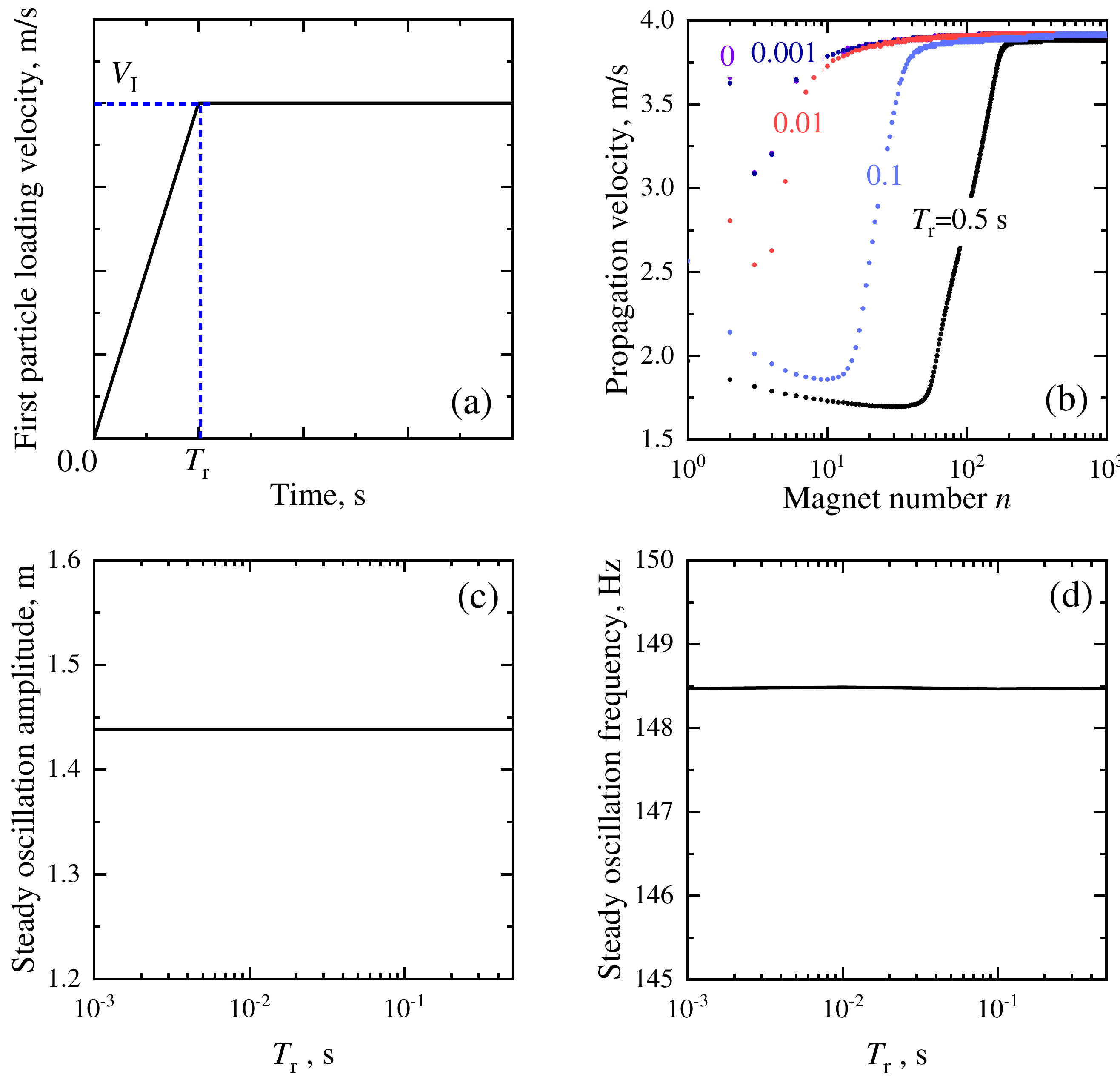}
  \caption{
(a) Schematic illustration of the ramp-constant velocity loading profile. (b) Dependence of propagation velocity on magnet number.  Steady oscillation amplitude (c) and oscillation frequency (d) verse ramp time \emph{T}\textsubscript{r}. Results are obtained by the calibrated simulation model with initial magnet separation $r_0=12$ mm and neglecting friction.}
 \label{fig.s7}
 \end{figure}

 Here we investigate the effect of the loading profile on the dynamic response of the nonlinear system. We first discuss the case characterized as a linear ramp - constant impact velocity (as shown in Fig. \ref{fig.s7}(a)). Fig. \ref{fig.s7}(b) shows that the loading profile with a longer acceleration time, $T_r$ (i.e. smaller acceleration) requires a longer propagation time to reach the steady propagation velocity, while the steady propagation velocity is independent of the acceleration time. Moreover, Figs. \ref{fig.s7}(c,d) show that the stabilized oscillation amplitude and frequency are also independent of the acceleration time.
 
\begin{figure}[ht!]
 \center
 \includegraphics[scale=0.36]{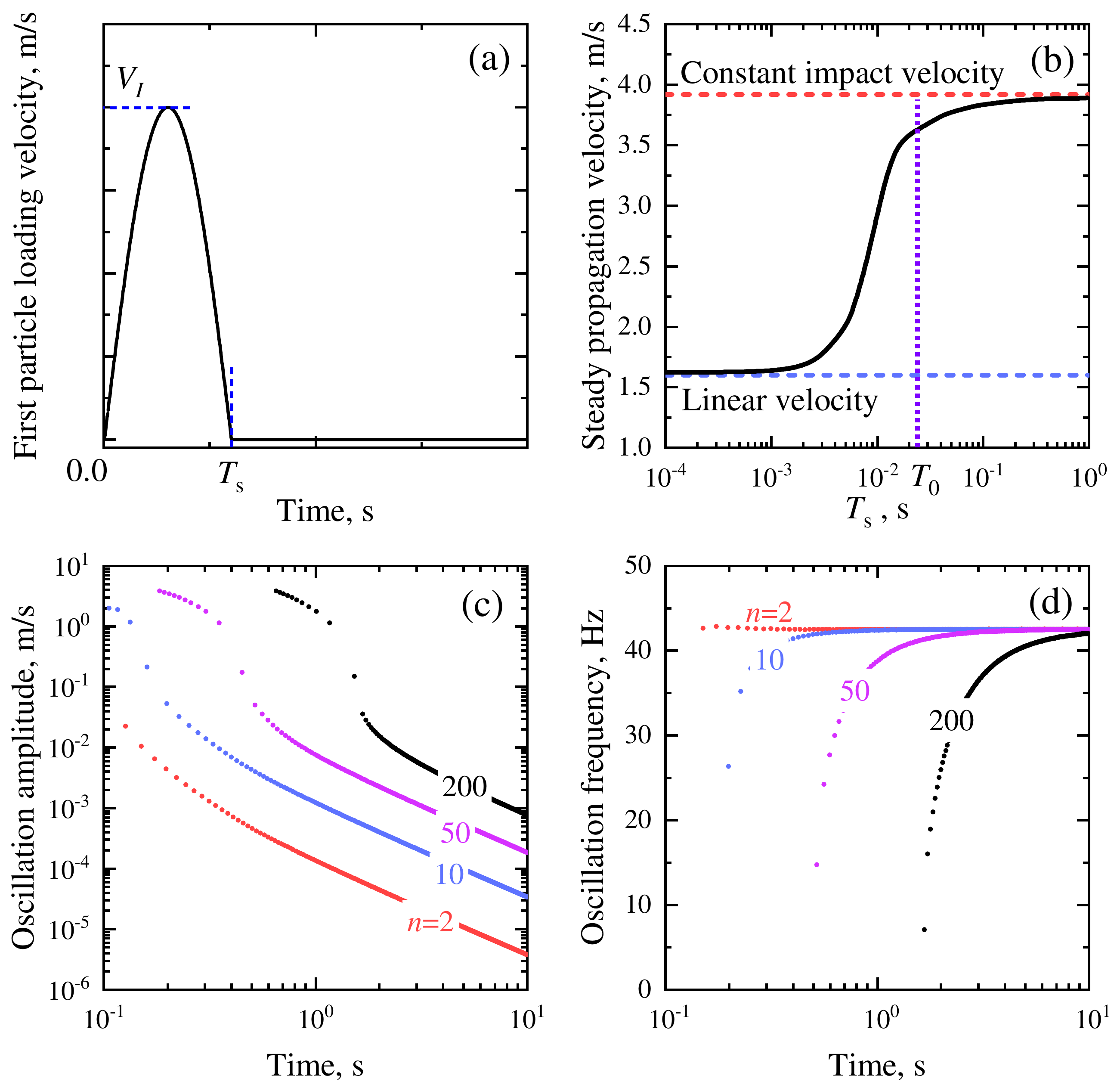}
  \caption{
(a) Schematic illustration of the velocity profile for a sinusoidal pulse impact wave. (b) Dependence of stabilized propagation velocity on signal duration time $T_s$, the peak impact velocity is set to be $V_I$=2 m/s. Oscillation amplitude (c) and oscillation frequency (d) verse propagation time, the signal duration time  $T_s$=0.1 s and $V_I$=2 m/s are considered. Results are obtained by the calibrated simulation model with initial magnet separation $r_0=12$ mm and neglecting friction.}
 \label{fig.s8}
 \end{figure}

 When considering a sinusoidal pulse impact wave (illustrated in Fig. \ref{fig.s8}a),  the steady propagation velocity decreases dramatically when the signal duration time is smaller than the linear oscillation period, $T_0$,  of the system (Fig. \ref{fig.s8}b) . Moreover, when $T_s<<T_0$, the steady propagation velocity tends to the linear propagation velocity; when $T_s>>T_0$ the steady propagation velocity approaches the propagation velocity of the case with constant impact velocity. After passing through the pulse wave,  Fig. \ref{fig.s8}(c,d) shows that the particle  oscillates rapidly decrease in both amplitude and period, until the amplitude finally vanishes, implying that the particles eventually reach a static equilibrium state.

  \vspace{50mm}
  \section{Effect of impactor velocity on the dynamic response of the nonlinear system}
\begin{figure}[ht!]
 \center
 \includegraphics[scale=0.4]{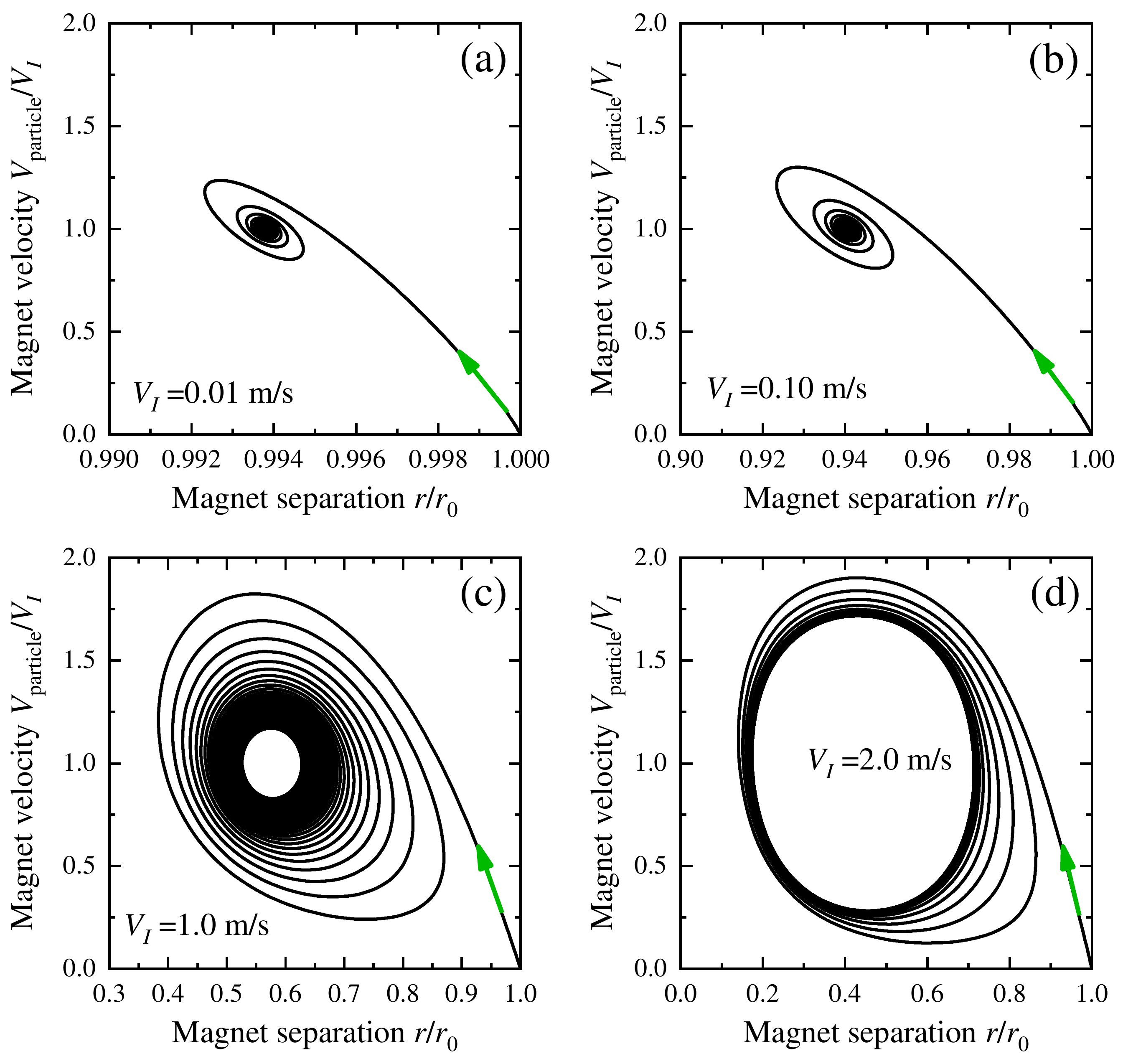}
  \caption{
Particle velocity versus magnet separation for various impactor velocities. (a) $V_I=0.01$  m/s, (b) $V_I=0.10$ m/s, (c) $V_I=1.0$ m/s, (d) $V_I=2.0$ m/s. Results are given for the $10^{\text{th}}$ magnet and are obtained by the calibrated simulation model with initial magnet separation $r_0=12$ mm and neglecting friction under constant impact velocity. The magnet separation is calculated as the gap between the $10^{\text{th}}$ and $11^{\text{th}}$ magnets.}
 \label{fig.s9}
 \end{figure}

  Here we show the effect of the impactor velocity on the dynamic response of the nonlinear system. Fig.\ref{fig.s9} shows the particle velocity as a function of magnet separation. Upon the arrival of the shock front, the particle velocity rapidly increases and oscillates about the impactor velocity. Remarkably, when moderate impactor velocity is considered, the particle finally reaches an steady state in which it adopts the velocity of impactor    (see Fig.\ref{fig.s9}(a) and (b)); however, as the impactor velocity exceeds a threshold value, the particle at long propagation time exhibits a steady oscillation about the impactor velocity, altering the particle kinetic energy and system potential energy (see Fig.\ref{fig.s9}(c) and (d)). Moreover, it is observed that the larger impactor velocity takes less transition cycles to reach the steady oscillatory state.

\end{document}